\documentclass[sn-mathphys-num]{sn-jnl}

\usepackage{bm,physics}
\usepackage{array}

\usepackage{graphicx}%
\usepackage{multirow}%
\usepackage{amsmath,amssymb,amsfonts}%
\usepackage{amsthm}%
\usepackage{mathrsfs}%
\usepackage[title]{appendix}%
\usepackage{xcolor}%
\usepackage{textcomp}%
\usepackage{manyfoot}%
\usepackage{booktabs}%
\usepackage{algorithm}%
\usepackage{algorithmicx}%
\usepackage{algpseudocode}%
\usepackage{listings}%

\begin{document}

\title[Faithfulness of Real-Space Renormalization Group Maps]{Faithfulness of Real-Space Renormalization Group Maps}

\author*[1]{\fnm{Katsuya O.} \sur{Akamatsu}}\email{akamatsu@issp.u-tokyo.ac.jp}

\author[1]{\fnm{Naoki} \sur{Kawashima}}\email{kawashima@issp.u-tokyo.ac.jp}

\affil*[1]{The Institute of Solid State Physics, University of Tokyo}

\abstract{The behavior of $b=2$ real-space renormalization group (RSRG) maps like the majority rule and the decimation map was examined by numerically applying RSRG steps to critical $q=2,3,4$ Potts spin configurations. While the majority rule is generally believed to work well, a more thorough investigation of the action of the map has yet to be considered in the literature. When fixing the size of the renormalized lattice $L_g$ and allowing the source configuration size $L_0$ to vary, we observed that the RG flow of the spin and energy correlation under the majority rule map appear to converge to a nontrivial model-dependent curve. We denote this property as ``faithfulness'', because it implies that some information remains preserved by RSRG maps that fall under this class. Furthermore, we show that $b=2$ weighted majority-like RSRG maps acting on the $q=2$ Potts model can be divided into two categories, maps that behave like decimation and maps that behave like the majority rule.}


\maketitle

\section{Introduction}

Real-space renormalization group methods like the Migdal-Kadanoff renormalization group scheme \cite{migdal1975,kadanoff1976} are used to probe the critical behavior of lattice spin models where each spin takes a finite number of states, operating directly in position space. This gives RSRG methods an intuitive and diagrammatic expression that make them particularly straightforward to grasp. However, the behavior of an RSRG method depends largely on the choice of map, and generally, these methods are difficult to analytically treat, with the notable exception of the decimation map, which can be exactly treated in 1D on an Ising chain. In 2D, the first step of the decimation map with a scale factor of $b=\sqrt{2}$ can be exactly computed, and for the case when $b=2$, the bond-moving approximation can be used to analytically treat the problem. \cite{wilson1975}

The majority rule map is another common choice of RSRG map. It has been used to study a variety of systems numerically, including, but not limited to, the ferromagnetic Ising model on triangular \cite{niemeijer1973}, square \cite{nauenberg1974}, and cubic lattices \cite{baillie1992}, and the Blume-Emery-Griffiths model on a triangular lattice \cite{adler1978}, among others. More recently, the majority rule has been used in machine learning studies that attempt to construct an optimal RG transformation \cite{chung2021} and an inverse RG map \cite{efthymiou2019}. Generally, it seems that there is consensus that the majority rule appears to work well, particularly for Ising-type systems. Analytically, it has been proven that the renormalized Hamiltonian under the $b=2$ majority rule remains well-defined slightly below the critical point for the square-lattice Ising model \cite{kennedy1993}. In this work, we numerically study the fixed-point behavior of the majority rule and decimation maps acting on Potts models, focusing on the RG flow generated by repeated applications of these maps on very large spin configurations.

In this study, we examine the $q$-state Potts model, where $q$ denotes the number of states, at the critical point, whose Hamiltonian is:
\begin{gather}
H=-\beta J\sum_{\expval{i\ j}}\delta_{s_is_j} \label{eq:potts_h}
\end{gather}
$\delta_{s_is_j}$ can be thought of as the Kronecker delta. More rigorously, it can be defined like \cite{wu1982}:
\begin{gather}
\delta_{s_is_j}=\frac{1}{q}\left[1+(q-1)(\hat{\bm{e}}_i\cdot\hat{\bm{e}}_j)\right] \label{eq:kronecker_def}
\end{gather}
The objects $\hat{\bm{e}}_i$ and $\hat{\bm{e}}_j$ are unit vectors located in a $(q-1)$-dimensional space. These unit vectors are taken from a set that exhibits the spin symmetry of the model, such that $\hat{\bm{e}}_i\cdot\hat{\bm{e}}_j=1$ when $i=j$, and $\hat{\bm{e}}_i\cdot\hat{\bm{e}}_j=-\frac{1}{q-1}$ otherwise. When $q=2$, the Potts model reduces to the Ising model. For $1\leq q\leq4$, the Potts model exhibits a continuous phase transition at the temperature \cite{beffara2012}:
\begin{gather}
T_c=\beta_c^{-1}=\ln\left[1+\sqrt{q}\right] \label{eq:tc}
\end{gather}
When $q>4$, the transition is first-order. In this work, we focus on the specific Potts models where $q\in\{2,3,4\}$, which correspond to instances of the Potts model for which the transition is second-order \cite{duminil-copin2017}.

We are particularly interested in the fixed-point behavior of RSRG maps when the renormalized lattice size $L_g$ is held to be fixed while increasing the source configuration size $L_0$. To this end, we introduce the property of faithfulness, which is operationally defined as the ability of a particular RSRG map to retain information about a critical source configuration in the limit of a large number of RSRG iterations $g$, fixing $L_g$. Under repeated applications of a faithful RG map, the distribution of configurations at a fixed $L_g$ tends towards a nontrivial fixed point distribution that shows the correct scaling behaviors. Furthermore, as $L_g$ increases, the estimates for the scaling exponents tend towards the correct values. On the other hand, an unfaithful RSRG map eventually loses all information about the critical source configuration (i.e. reaches a trivial fixed point) in the limit of large $g$. While the existence of a nontrivial fixed point under an RSRG map is independent of whether or not the resulting scaling behavior matches the original model, our definition of faithfulness is that a nontrivial fixed point exists and the correlation functions at the fixed point decay with the correct exponents.

More rigorously, we can consider a distribution function $P_L(S)$ of spin configurations $S$ on the system of size $L$ at criticality (with a certain boundary condition such as periodic boundary conditions). Then, an RG map is defined as a distribution function $P_L$ to $P'_{L/b}$. Let us consider the result of $g$ repeated applications of the same RG map, $P^{(g)}_{L_g}$, with $L_g\equiv Lb^{-g}$ and $P^{(0)}_{L_0}\equiv P_L$. By increasing $g$ while fixing the final system size to be $\Lambda \equiv L_g$, we can construct the series of distributions for the same system size, $P^{(0)}_\Lambda$, $P^{(1)}_\Lambda$, $P^{(2)}_\Lambda$, i.e., the distribution function $P^{(g)}_\Lambda$ is the result of $g$ repeated applications of the RG map to the original system of size $\Lambda b^g$. Now, we consider the ``thermodynamic'' limit of this series,
\begin{equation} \label{eq:thermo_limit}
P^*_\Lambda\equiv\lim_{g\to\infty}P^{(g)}_\Lambda
\end{equation}
We call the RG map is faithful when $P^*_\Lambda$ exists and possesses the same critical properties as the original distribution. Specifically, the correlation functions $\expval{\mathcal{O}_i\mathcal{O}_j}$ according to $P^*_\Lambda$ converges in the large $\Lambda$ limit for arbitrary local quantities $\mathcal{O}_i$ and $\mathcal{O}_j$, located at the positions, $i$ and $j$ respectively, shows the same asymptotic behavior, e.g., the same decay exponent for $P^*_\infty$ and $P_\infty$.

The decimation transformation is not faithful since the amplitude of the critical correlation function at the fixed normalized distance goes down to zero, that is:
\begin{equation} \label{eq:dec_corr}
\expval{s_0s_r}_g=\expval{s_0s_{rb^g}}_0\propto (b^g)^{-(d-2+\eta)}\to0,\quad g\to\infty
\end{equation}
Therefore, the distribution converges to a trivial distribution representing no correlation between spins. To obtain meaningful information from the decimation RG, the correlation function must be rescaled using the scaling dimension of the order parameter. In contrast, the majority rule provides us with the nontrivial fixed-point correlation function with no rescaling. Thus, it is a faithful map. In other words, the majority rule somehow ``knows'' the value of the scaling dimension. The faithfulness of the majority rule seems to have been taken for granted though there is no mathematical proof or convincing arguments in the literature, as far as we know.

\section{Methodology}

\subsection{Performing numerical RSRG}

The source critical configurations were generated using a parallelized cluster-flip Monte Carlo method. The details concerning the generated data are specified in Sec.~\ref{subsec:data}. To verify that the generated data was indeed critical, a Bayesian finite-size scaling analysis \cite{harada2011,harada2015} was performed on some sample output to check that the obtained critical exponents were close to the expected values.

We consider $b=2$ RSRG maps parametrized by a tensor $\bm{R}_{s';s_1,\cdots,s_4}$ whose elements satisfy:
\begin{gather}
0\leq\bm{R}_{s';s_1,\cdots,s_4}\leq1 \label{eq:r_elem1}\\
\sum_{s'}\bm{R}_{s';s_1,\cdots,s_4}=1 \label{eq:r_elem2}
\end{gather}
The variable $s'$ corresponds to the renormalized spin site on the new lattice, and the variables $s_1,\cdots,s_4$ corresponds to the spin sites within the block. In other words, Eqs.~\ref{eq:r_elem1} and \ref{eq:r_elem2} imply that the elements of $\bm{R}$ are the conditional probabilities that a particular value for the renormalized spin is selected to represent a block of spin values:
\begin{gather}
\bm{R}_{s';s_1,\cdots,s_4}=P(s'|s_1,\cdots,s_4) \label{eq:r_elem3}
\end{gather}
The RSRG map can then be repeatedly applied to a source configuration of size $L_0=b^{g_{max}}$ by iteratively constructing renormalized configurations based on the elements of $\bm{R}$ whose dimensions are reduced by a factor of $b$. A comprehensive discussion on RSRG maps treated as probability kernels can be found in \cite{vanenter1993}.

We can recover a deterministic RSRG map like $s'=T(s_1,\cdots,s_4)$ (mapping specific configurations of $s_1,\cdots,s_4$ to a renormalized spin $s'$) from the definition in Eq.~\ref{eq:r_elem3} by choosing a conditional probability like:
\begin{gather}
P(s'|s_1,\cdots,s_4)=\delta_{s',T(s_1,\cdots,s_4)} \label{eq:r_elem4}
\end{gather}

Because the lattices are finite, there is a maximum number of steps $g_{max}$ for which the map can be applied. While this approach is amenable to parallelization, it should be noted that the size of $\bm{R}$ grows exponentially like $\mathcal{O}(q^{b^2})$, with $q$ as the number of states that each spin can take. This means that explicitly specifying the elements of $\bm{R}$ is untenable for $b>2$, and an alternative approach such as storing the elements as a function could be used instead.

\subsection{Observables of interest and estimating the scaling exponents}\label{subsec:observables}

The observables of interest were the connected spin correlation function $G_g(r)$ and connected energy correlation function $\mathcal{G}_g(r)$, which were obtained from the two-point spin correlation function $\expval{\bm{s}_0\bm{s}_r}_g$ and the two-point local energy correlation function $\expval{\epsilon_0\epsilon_r}_g$ ($r$ denoting the distance between two spins and $g$ being the number of applied RSRG map iterations). Both quantities were exactly computed for each considered spin configuration. For the two-point spin correlation, the Potts spin state indices $s_i$ were interpreted as Potts spin vectors $\bm{s}_i$ in the simplex representation satisfying:
\begin{gather}
\bm{s}_i\cdot\bm{s}_j=\frac{q\delta_{ij}-1}{q-1} \label{eq:potts_simplex_dot}
\end{gather}
From the two-point spin correlation, the connected spin correlation function can be computed like:
\begin{gather}
G_g(r)=\expval{\overline{\bm{s}_0\bm{s}_r}}_g-\left|\expval{\overline{\bm{s}_0}}_g\right|^2 \label{eq:spin_corr}
\end{gather}
Here, the overline $\overline{\cdot}$ denotes a spatial average and the angle brackets $\expval{\cdot}$ denotes an average over configurations, which is essentially the thermal average since all Monte Carlo simulations were done at the same temperature. The second term is due to translational invariance for the Potts models considered, which implies $\expval{\overline{\bm{s}_0}}=\expval{\overline{\bm{s}_r}}$. Furthermore, the second term, which is associated with the magnetization, is expected to vanish on the infinite lattice at the critical point. The local energy at each site $i$ was defined to be:
\begin{gather}
\epsilon_i=\frac{1}{2}\sum_{j\in\expval{i\ j}}\delta_{s_is_j} \label{eq:local_energy}
\end{gather}
The sum runs over nearest-neighbor site pairs denoted by $\expval{i\ j}$. With this definition, the sum of the (negative) local energies at each site yields the energy of the configuration. The two-point energy correlation can then be computed by taking products of local energy operators at sites separated by distance $r$. Since the local energy is a scalar quantity, the connected energy correlation can be computed like:
\begin{gather}
\mathcal{G}_g(r)=\frac{\expval{\overline{\epsilon_0\epsilon_r}}_g-\expval{\overline{\epsilon_0}}_g^2}{\expval{\overline{\epsilon_0^2}}_g-\expval{\overline{\epsilon_0}}_g^2} \label{eq:energy_corr}
\end{gather}
Again, translational invariance was assumed in the above equations, so that $\expval{\overline{\epsilon_0}}=\expval{\overline{\epsilon_r}}$. The second term in the numerator converges to some non-zero constant on the infinite lattice, because the local energy as defined by Eq.~\ref{eq:local_energy} is nonnegative.

After computing the observables for each configuration, the performance of the RSRG map was gauged by checking the scaling behavior of the connected two-point functions $G_g(r)$ and $\mathcal{G}_g(r)$ across multiple iterations of the map. For critical configurations on infinite lattices, the connected spin correlation $G_g(r)$ is known to obey the following scaling:
\begin{gather}
G_g(r)\sim r^{2-d-\eta} \label{eq:spin_corr_dim}
\end{gather}
In Eq. \ref{eq:spin_corr_dim}, $d$ refers to the dimensionality ($d=2$ in this case) and $\eta=\Delta_{G(r)}=2\Delta_{s}$ is the related critical exponent, which is twice the dimension $\Delta_{s}$ of the spin operator in the CFT describing the model. This can be obtained from CFT considerations: the conformal weights $h_{u,v}$ of a particular operator can be described as a function of the central charge $c$ in the following manner \cite{francesco2012}:
\begin{gather}
h_{u,v}(c)=\frac{c-1}{24}+\frac{1}{4}\left(u\frac{\sqrt{1-c}+\sqrt{25-c}}{\sqrt{24}}+v\frac{\sqrt{1-c}-\sqrt{25-c}}{\sqrt{24}}\right)^2 \label{eq:cft_weights}
\end{gather}
This results in the following expression for the scaling dimension of the spin operator \cite{vasseur2014}:
\begin{gather}
\Delta_{s}=2h_{\frac{1}{2},0}=\frac{1}{96}\left(5+7c+\sqrt{(1-c)(25-c)}\right) \label{eq:spin_dim}
\end{gather}
As for the energy correlation, the theoretical value can be obtained by noting that the energy operator in minimal CFT models has a scaling dimension $\Delta_{\epsilon}$ equal to:
\begin{gather}
\Delta_{\epsilon}=2h_{2,1}=\frac{1}{8}\left(5-c+\sqrt{(1-c)(25-c)}\right) \label{eq:energy_dim}
\end{gather}
The dimension of the energy correlation is $\Delta_{\mathcal{G}(r)}=2\Delta_{\epsilon}$. The exact values for the scaling exponents of the spin and energy correlation in the Potts models considered were computed using Eqs.~\ref{eq:spin_dim} and \ref{eq:energy_dim} and are listed in Tab.~\ref{tab:scaling_dim}.
\begin{table}
\caption{Scaling dimensions for the spin and energy correlation from CFT considerations.}
\label{tab:scaling_dim}
\begin{tabular}{c|ccc}
\toprule
 & $c$ & $\Delta_{G(r)}=\eta$ & $\Delta_{\mathcal{G}(r)}$ \\
\midrule
$q=2$ & $\frac{1}{2}$ & $\frac{1}{4}$ & 2 \\
$q=3$ & $\frac{4}{5}$ & $\frac{4}{15}$ & $\frac{8}{5}$ \\
$q=4$ & 1 & $\frac{1}{4}$ & 1 \\
\bottomrule
\end{tabular}
\end{table}
The estimates for $G_g(r)$ (and $\mathcal{G}_g(r)$) were fit to a function of the form:
\begin{gather}
G_g(r)=c_1+\frac{c_2}{r^{c_3}} \label{eq:fit_func}
\end{gather}
In Eq.~\ref{eq:fit_func}, there are three free parameters: $c_1$ is the bias (which vanishes for the magnetization as the lattice size increases and is introduced to account for smaller lattice sizes), $c_2$ is some constant prefactor, and $c_3$ represents the estimated scaling dimension of interest (either $2\Delta_s$ or $2\Delta_\epsilon$). When $r$ is taken to be large, it is expected that the correlation function vanishes. Thus, at large $r$, it is possible that the noise exceeds the actual amplitude of the quantity of interest. Since the systems included periodic boundary conditions, the fit was done by excluding the two data points corresponding to the largest two values of $r$, so that the cutoff is at $r/4$ (since $b=2$).

Note that the data points obtained from the renormalized configurations are not independent. This is because the configurations are obtained by iterating an RSRG map over one set of source configurations, so that a renormalized configuration and its parent configuration are related.

\subsection{Types of RSRG maps considered}

\subsubsection{The majority rule and decimation maps for $b=2$}

The $b=2$ majority rule map and decimation map were the reference RSRG maps chosen for this study. The majority rule transformation describes the action of choosing the most frequent spin state amongst the spins in the block to represent the renormalized spin. In the Ising model, the spins only take two states $s_i=\pm1$, so that the majority rule can be defined by summing over the spins in the block and taking its sign. When there is a tie, the behavior can be defined to be a random choice between the two states. For the general case of Potts models, this approach is not easily generalizable, but can be done by instead counting the frequencies of each Potts state and taking the most frequent state. The RSRG tensor elements that define the $b=2$ majority rule map are:
\begin{gather}
\bm{R}_{s';s_1,\cdots,s_4}=
\begin{cases}
\frac{1}{||\text{mode}(s_1,\cdots,s_4)||},\quad s'\in\text{mode}(s_1,\cdots,s_4) \\
0,\quad s'\notin\text{mode}(s_1,\cdots,s_4)
\end{cases} \label{eq:maj}
\end{gather}
In Eq. \ref{eq:maj}, $\text{mode}(\cdots)$ refers to the set of statistical modes in the block, and $||\cdot||$ denotes the cardinality of the set. With this definition, when there is one mode (i.e. there is a clear majority), it is guaranteed to be the representative spin on the renormalized lattice, whereas when there are multiple modes, the renormalized spin is uniformly chosen from the modes.

On the other hand, the $b=2$ decimation map favoring the site $i_1$ is defined like:
\begin{gather}
\bm{R}_{s';s_1,\cdots,s_4}=
\begin{cases}
1,\quad s'=s_1 \\
0,\quad s'\neq s_1
\end{cases} \label{eq:dec}
\end{gather}
Here, Eq. \ref{eq:dec} guarantees that the spin at site $i_1$ is always chosen to be the renormalized spin. The decimation transformation is comparatively well-understood due to its relative simplicity. However, there is a known flaw in the decimation scheme where in the limit of a large number of decimation transformations on a configuration, the correlation function eventually decays to zero, which contradicts the expected behavior for a critical configuration \cite{wilson1975}.

\subsubsection{Weighted majority-like maps for $b=2$}

To interpolate between the majority rule and decimation maps, a parametrization using weights $w_j$ at each site $i_j$ was considered. The weights modify the relative contribution of the spins at each site, so that a weight of zero corresponds to having no influence on the choice of renormalized spin. Note that the weights are associated with the sites, and not with the different spin states. With this parametrization, the elements of the map for a weighted majority-like transformation are:
\begin{gather}
\bm{R}_{s';s_1,\cdots,s_4}=
\begin{cases}
\frac{1}{\left\lVert\underset{s\in\{s_k\}}{\mathrm{argmax}}\left\{\sum_{j=1}^{b^2}\delta_{s_j,s}w_j\right\}\right\rVert},\quad s'\in\underset{s\in\{s_k\}}{\mathrm{argmax}}\left\{\sum_{j=1}^{b^2}\delta_{s_j,s}w_j\right\} \\
0,\quad s'\notin\underset{s\in\{s_k\}}{\mathrm{argmax}}\left\{\sum_{j=1}^{b^2}\delta_{s_j,s}w_j\right\}
\end{cases} \label{eq:weighted_maj}
\end{gather}
From the definition in Eq. \ref{eq:weighted_maj}, it can be seen that the majority rule corresponds to the case when the weights are identical to each other, and the decimation map can be recovered when all but one weight is zero -- that is, only one spin influences the choice of the renormalized spin. Tab. \ref{tab:params} lists the parametrizations considered in this study.
\begin{table}
\caption{Parametrizations considered for the weighted majority-like maps with $b=2$. The abbreviation ``WM'' means ``weighted majority''. Our numerical results point to the faithfulness of the majority, WM1, and WM2 maps, and the unfaithfulness of the remaining maps.}
\label{tab:params}
\begin{tabular}{c|c}
\toprule
scheme & $(w_1,w_2,w_3,w_4)$\\
\midrule
majority & $(1,1,1,1)$\\
decimation & $(1,0,0,0)$\\
WM1 & $(1,0.5,0.5,0.5)$\\
WM2 & $(1,1,1,0)$\\
WM3 & $(1,0.5,0.25,0.125)$\\
WM4 & $(1,1,0,0)$\\
prob. majority & $(1,1,1,1)$\\
prob. decimation & $(1,0,0,0)$\\
prob. WM1 & $(1,0.5,0.5,0.5)$\\
prob. WM2 & $(1,1,1,0)$\\
prob. WM3 & $(1,0.5,0.25,0.125)$\\
prob. WM4 & $(1,1,0,0)$\\
\bottomrule
\end{tabular}
\end{table}

\subsubsection{Weighted probabilistic maps for $b=2$}

Another family of RSRG maps corresponding to probabalistic analogues of the weighted majority-like maps was also considered. For these maps, even when there is a clear majority, there is still a probability that the renormalized spin is chosen to be one of the spin states present in the block. The map is defined like:
\begin{gather}
\bm{R}_{s';s_1,\cdots,s_4}=\frac{\sum_{j=1}^{b^2}\delta_{s_j,s'}w_j}{\sum_{j=1}^{b^2}w_j} \label{eq:weighted_prob}
\end{gather}
The decimation map can still be obtained from Eq. \ref{eq:weighted_prob}, but this family of maps does not reduce to the majority rule map when all of the weights are identical.

\section{Results}

\subsection{Dataset details}\label{subsec:data}

The simulations were conducted using critical Potts configurations with $q\in\{2,3,4\}$ states and initial lattice sizes of $L_0\in\{256,512,1024,2048,4096\}$. For each combination of $q$ and $L_0$, 1600 samples were generated using a parallel Swendsen-Wang method. Each configuration was then repeatedly renormalized using the studied $b=2$ RSRG maps until the renormalized lattice size reached $L_{min}=16$, after which the observables associated with each configuration (initial or renormalized) were obtained.

\subsection{Numerical data on faithfulness}

\subsubsection{Behavior of the $b=2$ majority rule and decimation maps}
\begin{figure}
\centering
\includegraphics{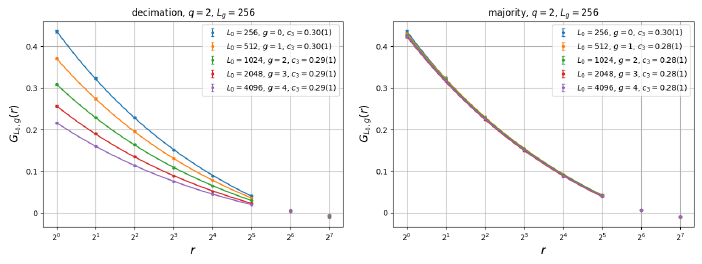}
\caption{Plots of the two-point spin correlation $G_g(r)$ of the $q=2$ Potts model as a function of the distance $r$, by the number of decimation (left) and majority rule (right) RSRG iterations $g$. Note that $L_g$ is held to be fixed to $L_g=256$. The expected power law has a scaling exponent of $2\Delta_s=\frac{1}{4}$.}
\label{fig:q2_potts_corr_lg_dec_maj}
\end{figure}
\begin{figure}
\centering
\includegraphics{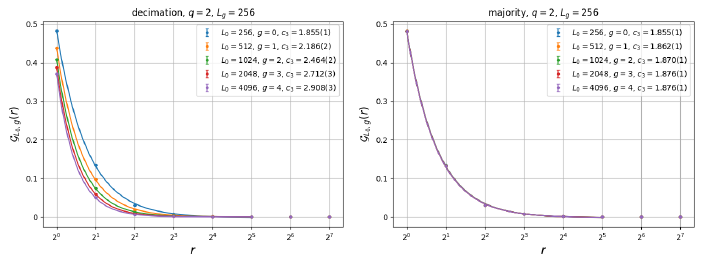}
\caption{Plots of the two-point energy correlation $\mathcal{G}_g(r)$ of the $q=2$ Potts model as a function of the distance $r$, by the number of decimation (left) and majority rule (right) RSRG iterations $g$. Note that $L_g$ is held to be fixed to $L_g=256$. The expected power law has a scaling exponent of $2\Delta_\epsilon=2$.}
\label{fig:q2_potts_ecorr_lg_dec_maj}
\end{figure}
Our numerical experiments show that the decimation and majority rule maps result in markedly different behavior when repeatedly applied to critical configurations. In what follows, we compare between the decimation map and the majority rule map. This is not for claiming the superiority of the majority-type rules to the decimation-type rules as a numerical technique for estimating the critical exponents, but rather, for establishing the majority rule's non-trivial property, i.e. its faithfulness, by contrasting the two types of mappings.

Fig.~\ref{fig:q2_potts_corr_lg_dec_maj} shows results for the $q=2$ (Ising) case, where, when fixing $L_g$, plots of the two-point spin correlation as a function of $r$ appear to lie very close to each other in the case of the majority rule map, but gradually shift downward with the decimation map. Adding to this, the two-point energy correlation shown in Fig.~\ref{fig:q2_potts_ecorr_lg_dec_maj} depicts a similar trend. In the case of decimation, the curves are shifted downwards as the number of applied steps $g$ increases (with fixed $L_g$) as expected. However, for the majority rule, we observed that the curves are all fairly close to each other (note the scale on the $y$-axis), and as $g$ increases, they appear to converge. These results serve as evidence for the faithfulness of the $b=2$ majority rule map. We observed very similar results for the $q=3$ and $q=4$ Potts models, for which the analogous plots are in the appendix.
\begin{figure}
\centering
\includegraphics{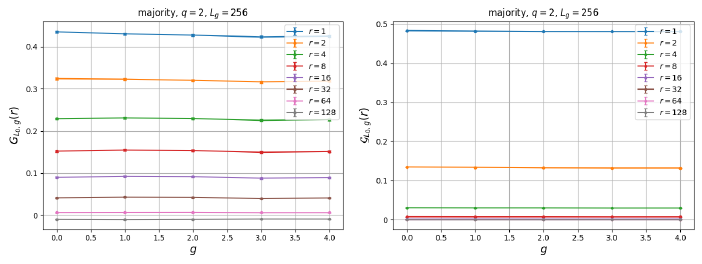}
\caption{Plots of the two-point spin correlation $G_g(r)$ (left) and energy correlation $\mathcal{G}_g(r)$ (right) of the $q=2$ Potts model as a function of the number of majority rule RSRG iterations $g$, by the distance $r$. Note that $L_g$ is held to be fixed to $L_g=256$. The lines serve as a guide for the eyes.}
\label{fig:q2_potts_lg_by_r_maj}
\end{figure}
The behavior of the majority rule is alternatively visualized in Fig.~\ref{fig:q2_potts_lg_by_r_maj}, where it is clear that repeated applications of the rule do not appear to change the two-point spin and energy correlation functions as $g$ increases. This is particularly evident for the spin correlation. For the energy correlation, the values decay initially, but appear to converge as $g$ increases. This is in contrast to the consistent decay in these observables for the decimation map.
\begin{figure}
\centering
\includegraphics{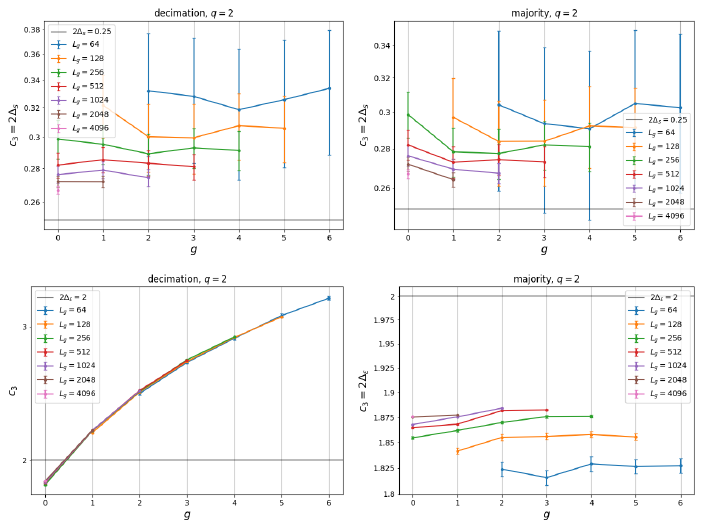}
\caption{Estimates of the coefficient $c_3$ as in Eq.~\ref{eq:fit_func}, which should agree with $2\Delta_s$ (top) and $2\Delta_\epsilon$ (bottom) provided that the mapping is faithful, for the $q=2$ Potts model as a function of the number of majority rule RSRG iterations $g$, by the renormalized lattice size $L_g$. The estimates were obtained from fitting according to the procedure outlined in Sec.~\ref{subsec:observables}, under the application of decimation (left) and majority rule (right). The solid gray lines indicate theoretical values.}
\label{fig:q2_potts_c_lg}
\end{figure}
By plotting the estimates (which were obtained as described in Sec.~\ref{subsec:observables}) for the scaling exponents $2\Delta_s$ and $2\Delta_\epsilon$ in Fig.~\ref{fig:q2_potts_c_lg}, it can be seen that both the decimation and majority rule maps produce estimates for $\Delta_s$ that converge to the expected value as $L_g$ gets larger and larger, staying relatively consistent as $g$ increases. This is expected behavior for the decimation transformation because it explicitly preserves the magnetization of the original configuration. For the majority rule, an argument for why it preserves the magnetization can be found in the appendix. However, the use of the decimation map does not result in an estimate for the energy correlation scaling exponent $\Delta_\epsilon$. This result can be visualized in Fig.~\ref{fig:q2_potts_c_lg} where the estimates for $c_3$ clearly do not converge to $\Delta_\epsilon$ under this map. On the other hand, however, the majority rule map produces estimates for $\Delta_\epsilon$ that tend towards the expected value as $L_g$ and $g$ increase, albeit slowly. This is further numerical evidence for the faithfulness of the majority rule map and similar maps, as well as the unfaithfulness of the decimation map and other maps that behave like it.

\subsubsection{Results for weighted majority-like maps}

In addition to observing the difference in the behavior of the decimation and majority rule maps, we also performed numerical experiments with the class of weighted majority-like rules defined by Eq.~\ref{eq:weighted_maj}. For this family of RSRG maps, we found that all of the specific parametrizations considered (Tab.~\ref{tab:params}) behaved in one of two ways. Some of the maps were found to behave like the decimation map, while others were found to behave like the majority rule map.
\begin{figure}
\centering
\includegraphics{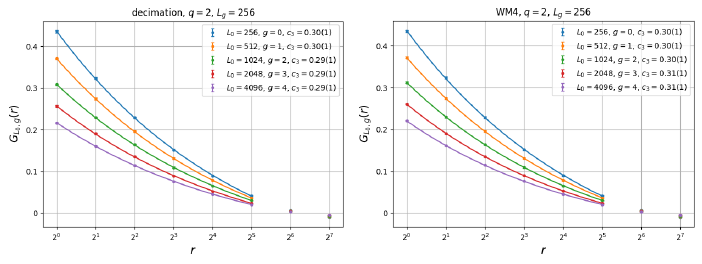}
\caption{Plots of the two-point spin correlation $G_g(r)$ of the $q=2$ Potts model as a function of the distance $r$, by the number of RSRG iterations $g$. The RSRG maps shown here behave like decimation, althought the ``WM3'' map was omitted since its action is identical to decimation. Note that $L_g$ is held to be fixed to $L_g=256$. The expected power law has a scaling exponent of $2\Delta_s=\frac{1}{4}$.}
\label{fig:q2_potts_corr_lg_dec_like}
\end{figure}
Fig.~\ref{fig:q2_potts_corr_lg_dec_like}, which collates plots of the spin correlation for some of the considered maps, demonstrates that some of the considered maps behave almost exactly like the $b=2$ decimation map. The same behavior was observed for the energy correlation for the relevant maps when compared to decimation. Note that the standard decimation map and the ``WM3'' case are also exactly identical. Furthermore, the ``WM1'' and ``WM2'' maps are identical to the majority rule map except for how ties are treated, and the same can be said for ``WM4'' and the decimation map.
\begin{figure}
\centering
\includegraphics{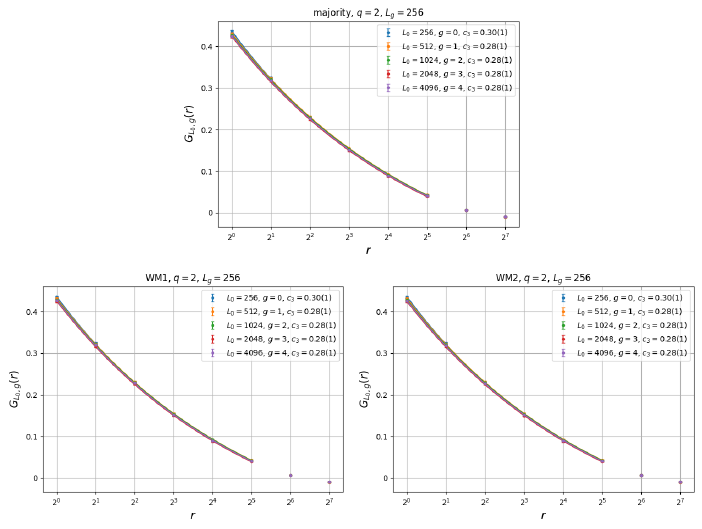}
\caption{Plots of the two-point spin correlation $G_g(r)$ of the $q=2$ Potts model as a function of the distance $r$, by the number of RSRG iterations $g$. The RSRG maps shown here behave like the majority rule. Note that $L_g$ is held to be fixed to $L_g=256$. The expected power law has a scaling exponent of $2\Delta_s=\frac{1}{4}$.}
\label{fig:q2_potts_corr_lg_maj_like}
\end{figure}
\begin{figure}
\centering
\includegraphics{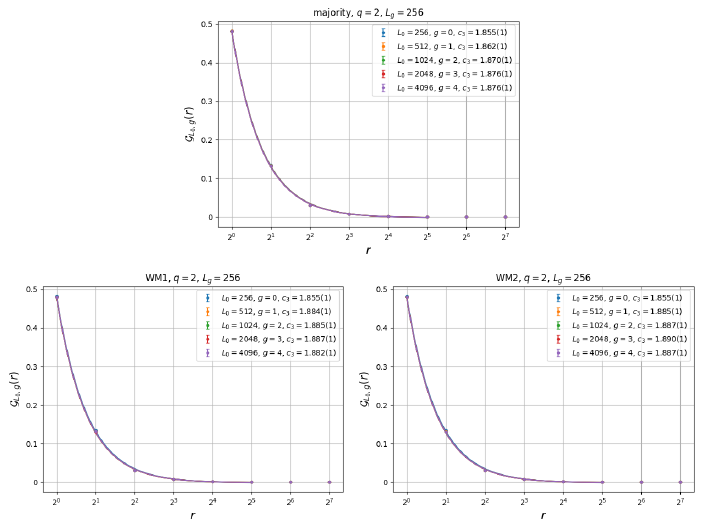}
\caption{Plots of the two-point energy correlation $\mathcal{G}_g(r)$ of the $q=2$ Potts model as a function of the distance $r$, by the number of RSRG iterations $g$. The RSRG maps shown here behave like the majority rule. Note that $L_g$ is held to be fixed to $L_g=256$. The expected power law has a scaling exponent of $2\Delta_\epsilon=2$.}
\label{fig:q2_potts_ecorr_lg_maj_like}
\end{figure}
Similarly, Figs.~\ref{fig:q2_potts_corr_lg_maj_like} and \ref{fig:q2_potts_ecorr_lg_maj_like} show maps that behave very similarly to the standard $b=2$ majority rule map. The plots suggest that unlike the decimation-like maps, the majority rule-like maps generate RG flows (fixing $L_g$, as usual) that do not appear to vanish in the limit of a large number of RSRG iterations $g$. Because the curves seem to converge to some limiting distribution as $g$ increases, it appears that these maps all fall in a broader class of RSRG maps that exhibit the property we denote as faithfulness. We also observed very similar results for the $q=3$ and $q=4$ Potts models.

\subsubsection{Results on probabilistic weighted maps}

\begin{figure}
\centering
\includegraphics{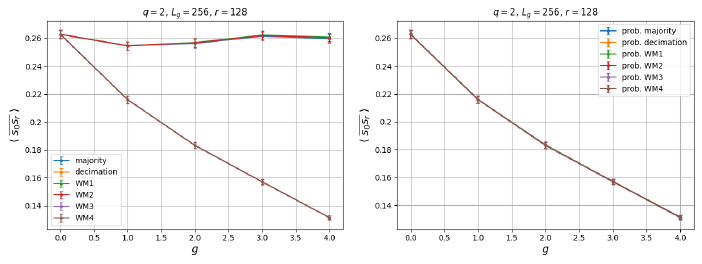}
\caption{Plots of the unconnected two-point spin correlation $\expval{\overline{s_0s_r}}$ of the $q=2$ Potts model as a function of the number of RG iterations $g$, by the RSRG map used. On the left are the deterministic maps, while the probabilistic maps are on the right. Note that $L_g$ is fixed to $L_g=256$ and $r$ is fixed to $r=128$.}
\label{fig:q2_potts_corr_lg_prob}
\end{figure}

Repeating the same analysis with the probabilistic weighted maps considered in Eq.~\ref{eq:weighted_prob} (on the $q=2,3,4$ Potts models, and using the parametrizations in Tab.~\ref{tab:params}) resulted in all considered maps, including the probabilistic analogue for the majority rule, behaving like the decimation transformation. This suggests that some degree of regularization (provided through the $\mathrm{argmax}$ function in Eq.~\ref{eq:weighted_maj}) is required to obtain behavior like the standard $b=2$ majority rule map. Otherwise, since these probabilistic weighted maps preserve the expectation value of the magnetization (without such regularization), the maps behave like the decimation transformation. The relevant plots for the $q=2$ Potts model are summarized in Fig.~\ref{fig:q2_potts_corr_lg_prob} for the spin correlation. It is clear from these plots that all of the probabilistic maps essentially behave like decimation (for decimation, the deterministic and probabilistic analogues are identical).

To see why all of these maps behave like decimation, we can consider a probabilistic map acting on some completely ordered region, and then imagine that a single spin is flipped by some fluctuation. If the RSRG map chooses this flipped spin as the representative spin for the block in the next iteration, the local fluctuation persists. The probabilistic map is thus always able to preserve or introduce fluctuations at every step, but for a faithful map, the probability of introducing a fluctuation must decrease as the number of RSRG steps increases. This is because introducing a fluctuation late into the RG procedure corresponds to a large fluctuation in the original model. Therefore, these maps cannot be faithful.

\subsection{Possible criteria for faithfulness}

A natural question that arises from the numerical observations is the question of what conditions are necessary in order for an RSRG map to behave faithfully. In particular, for the weighted majority-like maps, some of them behave like the majority rule, which appears to be faithful, whereas others behave like decimation, which is manifestly an unfaithful map. On the other hand, the probabilistic weighted maps all seem to behave like the standard decimation map.

Fortunately, Eq.~\ref{eq:weighted_maj} suggests that there are only a finite number of distinct $\bm{R}$ tensors representing weighted majority-like maps that can be constructed, following from the finiteness of the number of possible block configurations and the regularization used in the RSRG map. For $b=2$ weighted majority-like maps acting on critical configurations of the $q=2$ Potts model in particular, these $\bm{R}$ tensors can be exhaustively listed. Furthermore, the regions in parameter space that correspond to a specific realization of $\bm{R}$ can be mapped out.

\begin{figure}
\centering
\includegraphics{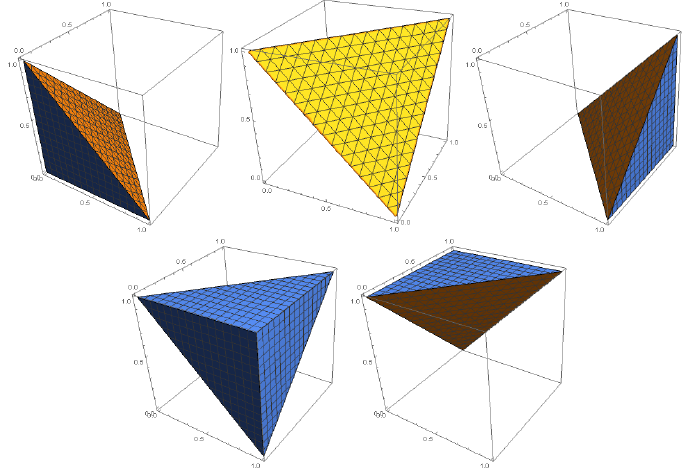}
\caption{Non-empty solution regions obtained from solving systems of the form given in Eq.~\ref{eq:regions_ineq_form_expanded}. Only the upper left region was found to behave like the decimation map.}
\label{fig:regions}
\end{figure}

To do this, we fixed one weight to be the largest weight, $w_1=1$, leaving three free parameters. Then, we considered systems of inequalities based on the form:
\begin{gather}
\begin{split}
w_1\sigma_1+w_2\sigma_2+w_3\sigma_3+w_4\sigma_4\lessgtr0 \\
\implies\pm1\pm w_2\pm w_3\pm w_4\lessgtr0 \label{eq:regions_ineq_form}
\end{split}
\end{gather}
While there appear to be 16 possibilities for the LHS in Eq.~\ref{eq:regions_ineq_form}, half of these choices are additive inverses of the other half. This leaves 8 choices for the LHS. We are interested in solving the system given by:
\begin{gather}
\begin{split}
1+w_2+w_3+w_4\lessgtr0,\quad1+w_2+w_3-w_4\lessgtr0 \\
1+w_2-w_3+w_4\lessgtr0,\quad1+w_2-w_3-w_4\lessgtr0 \\
1-w_2+w_3+w_4\lessgtr0,\quad1-w_2+w_3-w_4\lessgtr0 \\
1-w_2-w_3+w_4\lessgtr0,\quad1-w_2-w_3-w_4\lessgtr0
\end{split}
\label{eq:regions_ineq_form_expanded}
\end{gather}
There are $2^8=256$ instances of this system of inequalities to solve. Most of these systems result in no solution, but there are a few combinations for which solutions exist in the region of parameter space given by $w_2,w_3,w_4\in[0,1]$. There are five distinct solution regions that exist in the aforementioned space, which are shown in Fig.~\ref{fig:regions}. Of these five regions, only one region was found to result in decimation-like behavior, whereas the other four regions were found to behave like the majority rule map.
This region that describes the decimation-like maps, which is the first region in Fig.~\ref{fig:regions}, is described by the triangle-like inequality:
\begin{gather}
w_2+w_3+w_4\leq w_1 \label{eq:dec_region}
\end{gather}
The region that contains weighted majority-like maps that behave like the majority rule map (i.e. exhibit evidence of faithfulness) satisfy the inequality:
\begin{gather}
w_2+w_3+w_4>w_1 \label{eq:maj_region}
\end{gather}
We conjecture based on our numerical experiments that the same condition holds for the $q=3$ and $q=4$ Potts models, based on the numerical data that we have collected.

\section{Discussion}

Our main numerical results suggest that the majority rule map is a faithful map when applied to critical configurations of the $q$-state Potts models. We defined faithful RSRG maps to be maps that possess a nontrivial fixed point and correctly reproduce the expected scaling exponents. For the Ising model under the majority rule, we showed numerically that it possesses a nontrivial fixed point (the right panels of Figs.~\ref{fig:q2_potts_corr_lg_dec_maj} and \ref{fig:q2_potts_ecorr_lg_dec_maj}, or equivalently, Fig.~\ref{fig:q2_potts_lg_by_r_maj}). Furthermore, under the majority rule, the estimated coefficient $c_3$ approaches the expected scaling exponents (Fig.~\ref{fig:q2_potts_c_lg}) as the renormalized lattice size $L_g$ and number of RSRG iterations $g$ (and the source lattice size $L_0$) increase.

We found that the fixed point of the $b=2$ majority rule map on the Potts models appears to differ from the trivial fixed point approached by the $b=2$ decimation map. The majority rule seems to better preserve the information associated with the original configuration even after multiple iterations, and our results on this property are in line with those of another work \cite{lenggenhager2020} on the mutual information between source and renormalized Ising congfigurations under the decimation and majority rule maps. Note that the majority rule map is not necessarily the ``best'' map for estimating the critical exponents via a finite-size scaling analysis. In fact, RG transformations that attempt to improve on the majority rule map for the Ising model have been constructed in a previous work \cite{chung2021}.

Furthermore, we examined a broader family of RSRG maps in an attempt to clarify the mechanism behind the faithfulness of these maps. We found that considering weighted counts of spins in a block, which effectively constitutes a weighted majority-like rule, results in maps that behave like one of two distinct behaviors. These categories are prototyped by the majority rule and decimation map, and by extension, describe maps that are faithful and maps that are not faithful, respectively.

We also found that the regularization process employed by these maps, which was to rule out the possibility of assigning states that were not at the largest weighted count, was essential to obtaining a faithful map. The observation that all three studied models behaved similarly under the action of the examined maps suggests that the majority rule (and similar maps) is faithful when acting on systems exhibiting the $S_q$ Potts spin symmetry.

The faithful maps discussed in this paper have slight differences in their correlation profile, and we conjecture that faithful maps do not generate the same correlation profile when applied to the same model. In other words, the fixed-point correlation profile for a faithful RSRG map depends not only on the model (and on $L_g$), but also on the choice of map. In a similar vein, we conjecture that not all RSRG maps that exhibit nontrivial fixed-point behavior generate the correct exponents.

Future directions include examining the fixed-point behavior of other families of RSRG maps (in this direction, some results on $b=4$ weighted RSRG maps are presented in the appendix), and obtaining a clearer understanding of what characteristics a map must exhibit in order to be faithful.

\section*{Acknowledgements}
We would like to thank Satoshi Morita for insightful suggestions surrounding the development of the Monte Carlo code used in this study. We would also like to thank Xinliang Lyu and Kenji Homma for fruitful discussions. K.A. would like to acknowledge the support of the Global Science Graduate Course (GSGC) program of the University of Tokyo. The numerical calculations were done on nodes at the Supercomputer Center, the Institute of Solid State Physics, the University of Tokyo.

\begin{appendices}

\section{Numerical data for the $q=3$ and $q=4$ Potts models}

In this appendix, we show figures showing the behavior of the majority rule and decimation transformations on critical $q=3$ and $q=4$ Potts configurations.

\begin{figure}
\centering
\includegraphics{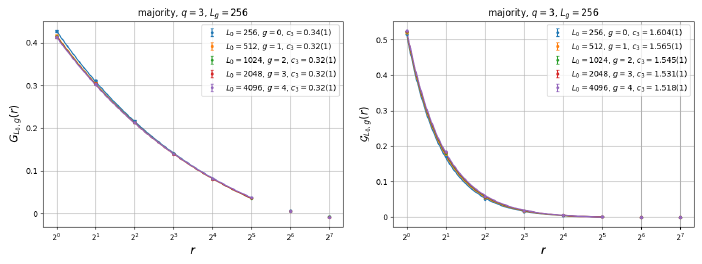}
\caption{Plots of the two-point spin correlation $G_g(r)$ and energy correlation $\mathcal{G}_g(r)$ of the $q=3$ Potts model as a function of the distance $r$, by the number of decimation (left) and majority rule (right) RSRG iterations $g$. Note that $L_g$ is held to be fixed to $L_g=256$. The expected power laws have scaling exponents of $2\Delta_s=\frac{4}{15}$ and $2\Delta_\epsilon=\frac{8}{5}$.}
\label{fig:q3_potts_corr_ecorr_lg_dec_maj}
\end{figure}

The plots for the $q=3$ Potts model (Fig.~\ref{fig:q3_potts_corr_ecorr_lg_dec_maj}) suggest the same interpretation as for the $q=2$ case. The data for the majority rule transformation appear to converge to some curve in the large-$g$ limit, when $L_g$ is fixed.
\begin{figure}
\centering
\includegraphics{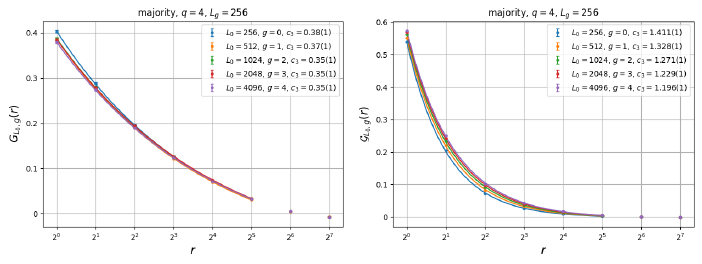}
\caption{Plots of the two-point spin correlation $G_g(r)$ and energy correlation $\mathcal{G}_g(r)$ of the $q=4$ Potts model as a function of the distance $r$, by the number of decimation (left) and majority rule (right) RSRG iterations $g$. Note that $L_g$ is held to be fixed to $L_g=256$. The expected power laws have scaling exponents of $2\Delta_s=\frac{1}{4}$ and $2\Delta_\epsilon=1$.}
\label{fig:q4_potts_corr_ecorr_lg_dec_maj}
\end{figure}
\begin{figure}
\centering
\includegraphics{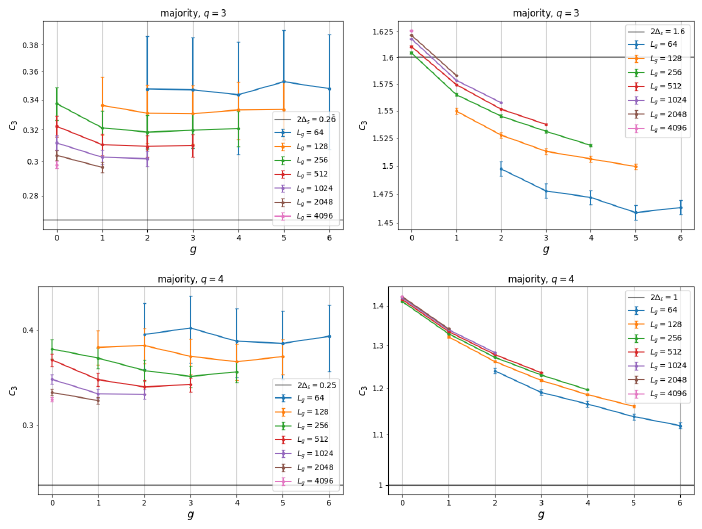}
\caption{Estimates of the coefficient $c_3$ as in Eq.~\ref{eq:fit_func}, which should agree with $2\Delta_s$ (left) and $2\Delta_\epsilon$ (right) provided that the mapping is faithful, for the $q=3$ (top) and $q=4$ (bottom) Potts models as a function of the number of majority rule RSRG iterations $g$, by the renormalized lattice size $L_g$. The estimates were obtained from fitting according to the procedure outlined in Sec.~\ref{subsec:observables}, under the application of decimation (left) and majority rule (right). The solid gray lines indicate theoretical values.}
\label{fig:q3_q4_potts_c_lg}
\end{figure}
Note that for the case of the $q=4$ Potts model, shown in Fig.~\ref{fig:q4_potts_corr_ecorr_lg_dec_maj}, the discrepancy between the expected power law using the known theoretical exponents and the data at $g=0$ may be due to the additional logarithmic correction present for the model. Furthermore, the figures for both the $q=3$ and $q=4$ models still demonstrate that the majority rule transformation acting on critical configurations possesses a nontrivial fized point, although more data at larger $L_g$ and $g$ may be needed to further examine the question of whether or not the correct scaling exponents are reproduced at large $L_g$ and $g$, particularly for the scaling dimension $\Delta_\epsilon$. From Fig.~\ref{fig:q3_q4_potts_c_lg}, the majority rule appears to at least preserve the exponent for the magnetization $\Delta_s$. We conjecture, for the Potts models under the majority rule map, that the estimate for $c_3$ when $L_g$ is large enough approaches the expected value for the scaling exponent when $g$ is large, for both the spin correlation and energy correlation (estimating $2\Delta_s$ and $2\Delta_\epsilon$, respectively).

Finally, while we omitted the results for the weighted maps and probabilistic analogues for the $q=3$ and $q=4$ Potts models for brevity, we found that the numerical data suggest similar conclusions as for the $q=2$ case.

\section{Correctness of the majority rule for the magnetization in the Ising model}

For the $q=2$ Potts model, which is identical to the Ising model, the majority rule map can be defined like:
\begin{gather}
s'_{i'}\equiv\text{sgn}(M_{i'})=\pm1 \label{eq:maj_ising_s} \\
M_{i'}\equiv b^{-d}\sum_{i\in\Lambda_b(i')}s_i \label{eq:maj_ising_m}
\end{gather}
In Eqs.~\ref{eq:maj_ising_s} and \ref{eq:maj_ising_m}, we consider a configuration $\bm{s}_i$ with Ising spins $s_i=\pm1$ on a $d$-dimensional hypercubic lattice. The primed variables $s',i'$ denote spins and sites on the renormalized lattice, respectively, and $\Lambda_b(i')$ refers to the block of spins in the preceding lattice that determines the spin at site $i'$. $M_{i'}$ is the magnetization associated with the block determining $s'_{i'}$. Our numerical data for the Ising model suggests that $\expval{\overline{s'_{i'}}}\propto b^{\Delta_s}$, where $\Delta_s$ is the scaling dimension for the spin and the expectation is taken over space and then over configurations. This is demonstrated in Fig.~\ref{fig:q2_potts_c_lg}, where the estimates for the scaling exponent approach $2\Delta_s$ as the renormalized lattice size $L_g$ increases. In other words, we found that the majority rule correctly preserves the magnetization of the original lattice.

In this section, we intend to demonstrate that application of the majority rule results in magnetization scaling of the form $\expval{\overline{s'_{i'}}}\propto b^{\Delta_s}$ as in the numerical observations. We start from the following standard scaling of the magnetization at the critical temperature:
\begin{gather}
m\sim L^{-\Delta_s} \label{eq:scaling_form_m}
\end{gather}
Here, $L$ represents the system size, $t$ is the rescaled temperature (where $t=0$ corresponds to the critical temperature), $y_t$ denotes the thermal exponent, and $\tilde{m}$ is a homogeneous function describing the scaling of the magnetization. Then, the block average of the magnetization $m'_{i'}$ for block size $b$ at the critical temperature scales like:
\begin{gather}
m'_{i'}\equiv b^{-d}M'_{i'}\propto b^{-\Delta_s} \label{eq:block_mag_scaling}
\end{gather}
This means that the distribution of block magnetizations at the critical temperature has a distribution function $P_b$ ($b$ denoting the scale factor) whose width is proportional to $b^{-\Delta_s}$. We then have the following scaling for the distribution function (using $\tilde{P}$ to denote the scaling function for $P_b$):
\begin{gather}
P_b(m')\sim b^{\Delta_s}\tilde{P}(m'b^{\Delta_s}) \label{eq:dist_scaling}
\end{gather}
Let us consider the expectation value $\expval{\overline{s'_{i'}}}$ for a typical configuration of a system with size $L$. Because of the fluctuation, the total magnetization of the whole system is not typically exactly zero. So, in what follows, without loss of generality, we assume that the magnetization is positive. For a system of size $L$, the magnetization per spin $m_0$ is typically on the order of $L^{-\Delta_s}$, which can be treated as the bias in estimating $\expval{\overline{s'_{i'}}}$. This has the effect of shifting the distribution for $m'(i')$ by $m_0$. The conditional probability for the block-average magnetization $m'$ given the average magnetization per spin $m_0$ for the whole system is:
\begin{gather}
P_b(m'|m_0)\sim P_b(m'-m_0)\sim b^{\Delta_s}\tilde{P}((m'-m_0)b^{\Delta_s}) \label{eq:shifted_dist}
\end{gather}
Then, we can calculate the expectation value of the renormalized spin $\expval{\overline{s'_{i'}}}$:
\begin{align}
\expval{\overline{s'_{i'}}}&\equiv \int_0^1dm'P_b(m'|m_0)-\int_{-1}^0dm'P_b(m'|m_0) \\
&\sim\int_{-m_0}^1dm'b^{\Delta_s}\tilde{P}(m'b^{\Delta_s})-\int_{-1}^{-m_0}dm'b^{\Delta_s}\tilde{P}(m'b^{\Delta_s}) \\
&=\int_{-m_0}^{m_0}dm'b^{\Delta_s}\tilde{P}(m'b^{\Delta_s})\sim2m_0b^{\Delta_s}\tilde{P}(0)\propto(bL^{-1})^{\Delta_s}
\end{align}
This shows that applying the majority rule map results in $b$-scaling that is consistent with our numerical observation on its correctness with regards to reproducing the scaling exponent of the magnetization, $\Delta_s$.

\section{Numerical data on the faithfulness of $b=4$ RSRG maps}

\begin{table}
\caption{Parametrizations considered for the weighted majority-like maps with $b=4$. The abbreviation ``WM'' means ``weighted majority''.}
\label{tab:params_b4}
\begin{tabular}{c|c}
\toprule
scheme & $w_i$\\
\midrule
majority & $w_1=\cdots=w_{16}=1$\\
decimation & $w_1=1,w_2=\cdots=w_{16}=0$\\
WM1-b4 & $w_1=1,w_2=\cdots=w_{16}=0.5$\\
WM2-b4 & $w_1=1,w_2=\cdots=w_{16}=0.07$\\
WM3-b4 & $w_1=1,w_2=\cdots=w_{16}=0.06$\\
\bottomrule
\end{tabular}
\end{table}

\begin{figure}
\centering
\includegraphics{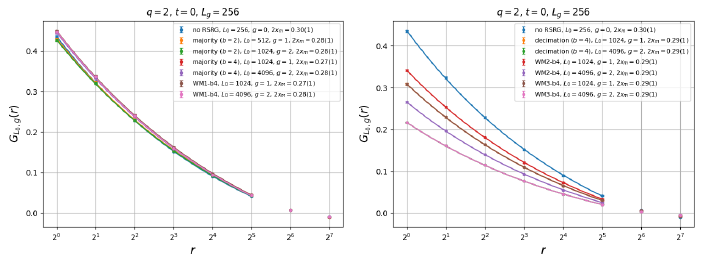}
\caption{Plots of the two-point spin correlation $G_g(r)$ of the $q=2$ Potts model as a function of the distance $r$, by the number of RSRG iterations $g$ and RSRG maps (left: faithful, right: unfaithful). The curves for the $b=4$ decimation map coincide with those of the ``WM3-b4'' map with the same $g$. Note that $L_g$ is held to be fixed to $L_g=256$. The expected power law has a scaling exponent of $2\Delta_s=\frac{1}{4}$, and was plotted using the prefactor and bias of the data at the lowest $g$.}
\label{fig:q2_potts_corr_lg_dec_maj_b4}
\end{figure}

\begin{figure}
\centering
\includegraphics{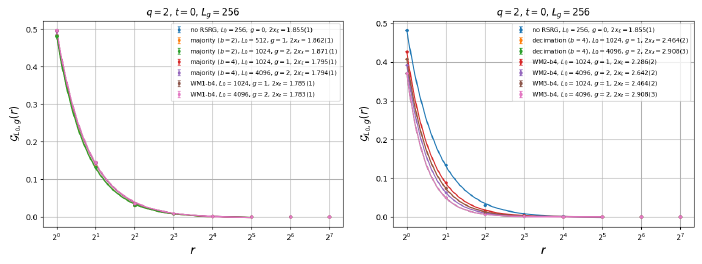}
\caption{Plots of the two-point spin correlation $G_g(r)$ of the $q=2$ Potts model as a function of the distance $r$, by the number of RSRG iterations $g$ and RSRG maps (left: faithful, right: unfaithful). The curves for the $b=4$ decimation map coincide with those of the ``WM3-b4'' map with the same $g$. Note that $L_g$ is held to be fixed to $L_g=256$. The expected power law has a scaling exponent of $2\Delta_\epsilon=2$, and was plotted using the prefactor and bias of the data at the lowest $g$. The curves associated with $4\times4$ decimation overlap with the curves associated with the WM3-b4 map.}
\label{fig:q2_potts_ecorr_lg_dec_maj_b4}
\end{figure}

In this section, we present some results for $b=4$ maps acting on critical configurations up to $L_0=4096$ and renormalized down to $L_{min}=16$. Since $b=4$, the weighted RSRG maps require specifying 16 weights. The parametrizations considered are listed in Tab.~\ref{tab:params_b4}, and they include parametrizations for the majority rule and decimation map on $4\times4$ blocks, weighted maps satisfying a $b=4$ version of Eq.~\ref{eq:maj_region} (WM1-b4, which easily satisfies the inequality, and WM2-b4, which is close to saturating the inequality), and a weighted map satisfying a $b=4$ version of Eq.~\ref{eq:dec_region} (WM3-b4, for which $w_2,...,w_{16}$ sum to slightly less than $w_1$). The corresponding versions of Eq.~\ref{eq:maj_region} and Eq.~\ref{eq:dec_region} are:
\begin{gather}
\sum_{i=2}^{16}w_i>w_1 \label{eq:maj_region_b4} \\
\sum_{i=2}^{16}w_i\leq w_1 \label{eq:dec_region_b4}
\end{gather}
The results, presented in Fig.~\ref{fig:q2_potts_corr_lg_dec_maj_b4} and Fig.~\ref{fig:q2_potts_ecorr_lg_dec_maj_b4}, indicate that the majority rule and WM1-b4 maps are faithful maps, while the WM2-b4 and WM3-b4 maps are unfaithful and behave like decimation. In particular, WM3-b4, which satisfies Eq.~\ref{eq:dec_region_b4}, reduces to decimation and generates identical behavior, while WM2-b4, which satisfies Eq.~\ref{eq:maj_region_b4}, does not behave like the $4\times4$ majority rule map and instead behaves qualitatively like the $4\times4$ decimation map. The WM2-b4 map produces a different profile from the decimation and WM3-b4 maps. The behavior of the WM2-b4 map suggests that a naive generalization of Eqs.~\ref{eq:dec_region} and ~\ref{eq:maj_region} is insufficient to distinguish between faithful and unfaithful maps.
The $b=4$ faithful maps behave slightly differently compared to the case of the $b=2$ majority rule, but these $b=4$ maps are still able to produce estimates for the critical exponents $\Delta_s$ and $\Delta_\epsilon$ similar to those of the $b=2$ faithful maps. On the other hand, the unfaithful $b=4$ maps produce estimates for $\Delta_\epsilon$ that get worse as $L_g$ and $g$ increase.

\end{appendices}

\bibliography{sn-bibliography}


\begin{thebibliography}{19}
\ifx \bisbn   \undefined \def \bisbn  #1{ISBN #1}\fi
\ifx \binits  \undefined \def \binits#1{#1}\fi
\ifx \bauthor  \undefined \def \bauthor#1{#1}\fi
\ifx \batitle  \undefined \def \batitle#1{#1}\fi
\ifx \bjtitle  \undefined \def \bjtitle#1{#1}\fi
\ifx \bvolume  \undefined \def \bvolume#1{\textbf{#1}}\fi
\ifx \byear  \undefined \def \byear#1{#1}\fi
\ifx \bissue  \undefined \def \bissue#1{#1}\fi
\ifx \bfpage  \undefined \def \bfpage#1{#1}\fi
\ifx \blpage  \undefined \def \blpage #1{#1}\fi
\ifx \burl  \undefined \def \burl#1{\textsf{#1}}\fi
\ifx \doiurl  \undefined \def \doiurl#1{\url{https://doi.org/#1}}\fi
\ifx \betal  \undefined \def \betal{\textit{et al.}}\fi
\ifx \binstitute  \undefined \def \binstitute#1{#1}\fi
\ifx \binstitutionaled  \undefined \def \binstitutionaled#1{#1}\fi
\ifx \bctitle  \undefined \def \bctitle#1{#1}\fi
\ifx \beditor  \undefined \def \beditor#1{#1}\fi
\ifx \bpublisher  \undefined \def \bpublisher#1{#1}\fi
\ifx \bbtitle  \undefined \def \bbtitle#1{#1}\fi
\ifx \bedition  \undefined \def \bedition#1{#1}\fi
\ifx \bseriesno  \undefined \def \bseriesno#1{#1}\fi
\ifx \blocation  \undefined \def \blocation#1{#1}\fi
\ifx \bsertitle  \undefined \def \bsertitle#1{#1}\fi
\ifx \bsnm \undefined \def \bsnm#1{#1}\fi
\ifx \bsuffix \undefined \def \bsuffix#1{#1}\fi
\ifx \bparticle \undefined \def \bparticle#1{#1}\fi
\ifx \barticle \undefined \def \barticle#1{#1}\fi
\bibcommenthead
\ifx \bconfdate \undefined \def \bconfdate #1{#1}\fi
\ifx \botherref \undefined \def \botherref #1{#1}\fi
\ifx \url \undefined \def \url#1{\textsf{#1}}\fi
\ifx \bchapter \undefined \def \bchapter#1{#1}\fi
\ifx \bbook \undefined \def \bbook#1{#1}\fi
\ifx \bcomment \undefined \def \bcomment#1{#1}\fi
\ifx \oauthor \undefined \def \oauthor#1{#1}\fi
\ifx \citeauthoryear \undefined \def \citeauthoryear#1{#1}\fi
\ifx \endbibitem  \undefined \def \endbibitem {}\fi
\ifx \bconflocation  \undefined \def \bconflocation#1{#1}\fi
\ifx \arxivurl  \undefined \def \arxivurl#1{\textsf{#1}}\fi
\csname PreBibitemsHook\endcsname

\bibitem[\protect\citeauthoryear{Migdal}{1975}]{migdal1975}
\begin{barticle}
\bauthor{\bsnm{Migdal}, \binits{A.}}:
\batitle{Phase transitions in gauge and spin-lattice systems}.
\bjtitle{Zh. Eksp. Teor. Fiz}
\bvolume{69},
\bfpage{1457}
(\byear{1975})
\end{barticle}
\endbibitem

\bibitem[\protect\citeauthoryear{Kadanoff}{1976}]{kadanoff1976}
\begin{barticle}
\bauthor{\bsnm{Kadanoff}, \binits{L.P.}}:
\batitle{Notes on {M}igdal's recursion formulas}.
\bjtitle{Annals of Physics}
\bvolume{100}(\bissue{1-2}),
\bfpage{359}--\blpage{394}
(\byear{1976})
\end{barticle}
\endbibitem

\bibitem[\protect\citeauthoryear{Wilson}{1975}]{wilson1975}
\begin{barticle}
\bauthor{\bsnm{Wilson}, \binits{K.G.}}:
\batitle{The renormalization group: {C}ritical phenomena and the {K}ondo
  problem}.
\bjtitle{Reviews of modern physics}
\bvolume{47}(\bissue{4}),
\bfpage{773}
(\byear{1975})
\end{barticle}
\endbibitem

\bibitem[\protect\citeauthoryear{Niemeijer and
  Van~Leeuwen}{1973}]{niemeijer1973}
\begin{barticle}
\bauthor{\bsnm{Niemeijer}, \binits{T.}},
\bauthor{\bsnm{Van~Leeuwen}, \binits{J.}}:
\batitle{Wilson theory for spin systems on a triangular lattice}.
\bjtitle{Physical Review Letters}
\bvolume{31}(\bissue{23}),
\bfpage{1411}
(\byear{1973})
\end{barticle}
\endbibitem

\bibitem[\protect\citeauthoryear{Nauenberg and Nienhuis}{1974}]{nauenberg1974}
\begin{barticle}
\bauthor{\bsnm{Nauenberg}, \binits{M.}},
\bauthor{\bsnm{Nienhuis}, \binits{B.}}:
\batitle{Critical surface for square {I}sing spin lattice}.
\bjtitle{Physical Review Letters}
\bvolume{33}(\bissue{16}),
\bfpage{944}
(\byear{1974})
\end{barticle}
\endbibitem

\bibitem[\protect\citeauthoryear{Baillie et~al.}{1992}]{baillie1992}
\begin{barticle}
\bauthor{\bsnm{Baillie}, \binits{C.F.}},
\bauthor{\bsnm{Gupta}, \binits{R.}},
\bauthor{\bsnm{Hawick}, \binits{K.A.}},
\bauthor{\bsnm{Pawley}, \binits{G.S.}}:
\batitle{Monte {C}arlo renormalization-group study of the three-dimensional
  ising model}.
\bjtitle{Physical Review B}
\bvolume{45}(\bissue{18}),
\bfpage{10438}
(\byear{1992})
\end{barticle}
\endbibitem

\bibitem[\protect\citeauthoryear{Adler et~al.}{1978}]{adler1978}
\begin{barticle}
\bauthor{\bsnm{Adler}, \binits{J.}},
\bauthor{\bsnm{Aharony}, \binits{A.}},
\bauthor{\bsnm{Oitmaa}, \binits{J.}}:
\batitle{Renormalisation group studies of the {B}lume-{E}mery-{G}riffiths model
  in two dimensions}.
\bjtitle{Journal of Physics A: Mathematical and General}
\bvolume{11}(\bissue{5}),
\bfpage{963}
(\byear{1978})
\end{barticle}
\endbibitem

\bibitem[\protect\citeauthoryear{Chung and Kao}{2021}]{chung2021}
\begin{barticle}
\bauthor{\bsnm{Chung}, \binits{J.-H.}},
\bauthor{\bsnm{Kao}, \binits{Y.-J.}}:
\batitle{Neural {M}onte {C}arlo renormalization group}.
\bjtitle{Physical Review Research}
\bvolume{3}(\bissue{2}),
\bfpage{023230}
(\byear{2021})
\end{barticle}
\endbibitem

\bibitem[\protect\citeauthoryear{Efthymiou et~al.}{2019}]{efthymiou2019}
\begin{barticle}
\bauthor{\bsnm{Efthymiou}, \binits{S.}},
\bauthor{\bsnm{Beach}, \binits{M.J.}},
\bauthor{\bsnm{Melko}, \binits{R.G.}}:
\batitle{Super-resolving the {I}sing model with convolutional neural networks}.
\bjtitle{Physical Review B}
\bvolume{99}(\bissue{7}),
\bfpage{075113}
(\byear{2019})
\end{barticle}
\endbibitem

\bibitem[\protect\citeauthoryear{Kennedy}{1993}]{kennedy1993}
\begin{barticle}
\bauthor{\bsnm{Kennedy}, \binits{T.}}:
\batitle{Some rigorous results on majority rule renormalization group
  transformations near the critical point}.
\bjtitle{Journal of Statistical Physics}
\bvolume{72},
\bfpage{15}--\blpage{37}
(\byear{1993})
\end{barticle}
\endbibitem

\bibitem[\protect\citeauthoryear{Wu}{1982}]{wu1982}
\begin{barticle}
\bauthor{\bsnm{Wu}, \binits{F.-Y.}}:
\batitle{The {P}otts model}.
\bjtitle{Reviews of Modern Physics}
\bvolume{54}(\bissue{1}),
\bfpage{235}
(\byear{1982})
\end{barticle}
\endbibitem

\bibitem[\protect\citeauthoryear{Beffara and Duminil-Copin}{2012}]{beffara2012}
\begin{barticle}
\bauthor{\bsnm{Beffara}, \binits{V.}},
\bauthor{\bsnm{Duminil-Copin}, \binits{H.}}:
\batitle{The self-dual point of the two-dimensional random-cluster model is
  critical for $q\geq1$}.
\bjtitle{Probability Theory and Related Fields}
\bvolume{153}(\bissue{3-4}),
\bfpage{511}--\blpage{542}
(\byear{2012})
\end{barticle}
\endbibitem

\bibitem[\protect\citeauthoryear{Duminil-Copin
  et~al.}{2017}]{duminil-copin2017}
\begin{barticle}
\bauthor{\bsnm{Duminil-Copin}, \binits{H.}},
\bauthor{\bsnm{Sidoravicius}, \binits{V.}},
\bauthor{\bsnm{Tassion}, \binits{V.}}:
\batitle{Continuity of the phase transition for planar random-cluster and potts
  models with $1\leq q\leq4$}.
\bjtitle{Communications in Mathematical Physics}
\bvolume{349},
\bfpage{47}--\blpage{107}
(\byear{2017})
\end{barticle}
\endbibitem

\bibitem[\protect\citeauthoryear{Harada}{2011}]{harada2011}
\begin{barticle}
\bauthor{\bsnm{Harada}, \binits{K.}}:
\batitle{Bayesian inference in the scaling analysis of critical phenomena}.
\bjtitle{Physical Review E}
\bvolume{84}(\bissue{5}),
\bfpage{056704}
(\byear{2011})
\end{barticle}
\endbibitem

\bibitem[\protect\citeauthoryear{Harada}{2015}]{harada2015}
\begin{barticle}
\bauthor{\bsnm{Harada}, \binits{K.}}:
\batitle{Kernel method for corrections to scaling}.
\bjtitle{Physical Review E}
\bvolume{92}(\bissue{1}),
\bfpage{012106}
(\byear{2015})
\end{barticle}
\endbibitem

\bibitem[\protect\citeauthoryear{Van~Enter et~al.}{1993}]{vanenter1993}
\begin{barticle}
\bauthor{\bsnm{Van~Enter}, \binits{A.C.}},
\bauthor{\bsnm{Fern{\'a}ndez}, \binits{R.}},
\bauthor{\bsnm{Sokal}, \binits{A.D.}}:
\batitle{Regularity properties and pathologies of position-space
  renormalization-group transformations: {S}cope and limitations of {G}ibbsian
  theory}.
\bjtitle{Journal of Statistical Physics}
\bvolume{72},
\bfpage{879}--\blpage{1167}
(\byear{1993})
\end{barticle}
\endbibitem

\bibitem[\protect\citeauthoryear{Francesco et~al.}{2012}]{francesco2012}
\begin{bbook}
\bauthor{\bsnm{Francesco}, \binits{P.}},
\bauthor{\bsnm{Mathieu}, \binits{P.}},
\bauthor{\bsnm{S{\'e}n{\'e}chal}, \binits{D.}}:
\bbtitle{Conformal Field Theory}.
\bpublisher{Springer},
\blocation{Berlin}
(\byear{2012})
\end{bbook}
\endbibitem

\bibitem[\protect\citeauthoryear{Vasseur and Jacobsen}{2014}]{vasseur2014}
\begin{barticle}
\bauthor{\bsnm{Vasseur}, \binits{R.}},
\bauthor{\bsnm{Jacobsen}, \binits{J.L.}}:
\batitle{Operator content of the critical {P}otts model in $d$ dimensions and
  logarithmic correlations}.
\bjtitle{Nuclear Physics B}
\bvolume{880},
\bfpage{435}--\blpage{475}
(\byear{2014})
\end{barticle}
\endbibitem

\bibitem[\protect\citeauthoryear{Lenggenhager et~al.}{2020}]{lenggenhager2020}
\begin{barticle}
\bauthor{\bsnm{Lenggenhager}, \binits{P.M.}},
\bauthor{\bsnm{G{\"o}kmen}, \binits{D.E.}},
\bauthor{\bsnm{Ringel}, \binits{Z.}},
\bauthor{\bsnm{Huber}, \binits{S.D.}},
\bauthor{\bsnm{Koch-Janusz}, \binits{M.}}:
\batitle{Optimal renormalization group transformation from information theory}.
\bjtitle{Physical Review X}
\bvolume{10}(\bissue{1}),
\bfpage{011037}
(\byear{2020})
\end{barticle}
\endbibitem

\end{thebibliography}

\end{document}